\documentstyle[12pt,epsf]{article}
\newlength{\extraspace}
\newlength{\extraspaces}
\newcommand{\be}{\begin{equation}
\addtolength{\abovedisplayskip}{\extraspaces}
\addtolength{\belowdisplayskip}{\extraspaces}
\addtolength{\abovedisplayshortskip}{\extraspace}
\addtolength{\belowdisplayshortskip}{\extraspace}}
\newcommand{\ee}{\end{equation}}
 
\newcommand{\ba}{\begin{eqnarray}
\addtolength{\abovedisplayskip}{\extraspaces}
\addtolength{\belowdisplayskip}{\extraspaces}
\addtolength{\abovedisplayshortskip}{\extraspace}
\addtolength{\belowdisplayshortskip}{\extraspace}}
\newcommand{\ea}{\end{eqnarray}}

\newcommand{\bas}{\begin{eqnarray*}
\addtolength{\abovedisplayskip}{\extraspaces}
\addtolength{\belowdisplayskip}{\extraspaces}
\addtolength{\abovedisplayshortskip}{\extraspace}
\addtolength{\belowdisplayshortskip}{\extraspace}}
\newcommand{\eas}{\end{eqnarray*}}
 
\newcounter{subequation}[equation]
\makeatletter

\expandafter\let\expandafter
\reset@font\csname reset@font\endcsname

\def\subeqnarray{\arraycolsep1pt
    \def\@eqnnum\stepcounter##1{\stepcounter{subequation}%
        {\reset@font\rm(\theequation\alph{subequation})}}
\jot5mm     \eqnarray}

\makeatother

\newcommand{\newsection}[1]{
\vspace{4mm}
\pagebreak[3]
\addtocounter{section}{1}
 
\begin{flushleft}
{\large\bf \thesection. #1}
\end{flushleft}
\nopagebreak
\vspace{-3mm}

\nopagebreak}

\newcommand{\NP}[1]{Nucl.\ Phys.\ {\bf #1}}
\newcommand{\PL}[1]{Phys.\ Lett.\ {\bf #1}}

\newcommand{\IJMP}[1]{Int.\ J.\ Mod.\ Phys.\ {\bf #1}}

\newcommand{\N}{\mbox{I\hspace{-.4ex}N}}
\newcommand{\C}{\mbox{$\,${\sf I}\hspace{-1.2ex}{\bf C}}}

\newcommand{\R}{\mbox{\rm I\hspace{-.4ex}R}}

\newcommand{\bra}{\langle}
\newcommand{\ket}{\rangle}

\newcommand{\rra}{\ \longrightarrow \ }

\newcommand{\is}{ &\!=\!& }
\newcommand{\nonum}{\nonumber \\[1.5mm]}
\newcommand{\sspace}{\makebox[1cm]{ }}

\newcommand{\lsim}{ \stackrel{\textstyle{<}}{\sim} }
\newcommand{\gsim}{ \stackrel{\textstyle{>}}{\sim} }

\newcommand{\th}{{\theta}}

\newcommand{\sh}{{\rm sh}}
\newcommand{\ch}{{\rm ch}}
\renewcommand{\d}{{\partial}}

\newcommand{\Gbar}{{\overline{G}}}

\newcommand{\gbar}{{\overline{g}}}

\newcommand{\cO}{{\cal O}}

\begin{document}
%
\begin{titlepage}
%
\renewcommand{\thefootnote}{\fnsymbol{footnote}}
\begin{flushright}
MPI-PhT/96-25\\
hep-th/9604161
\end{flushright}
\vspace{1cm}

\begin{center}
{\LARGE Perturbative versus Non-perturbative QFT\\[3mm]
Lessons from the $O(3)$ NLS Model}
\vspace{2cm}
 
{\large J. Balog\footnote{On leave of absence from the Central
Research Institute for Physics, Budapest, Hungary},
M. Niedermaier}\\ [3mm]
{\small\sl Max-Planck-Institut f\"{u}r Physik}\\
{-\small\sl Werner Heisenberg Institut - }\\
{\small\sl F\"{o}hringer Ring 6, 80805 Munich, Germany}
\vspace{0.5cm}

{\large T. Hauer}\\ [3mm]
{\small\sl Center for Theoretical Physics, Laboratory for Nuclear Science}\\
{\small\sl Massachusetts Institute of Technology}\\
{\small\sl Cambridge, MA 02139, USA} 

\vspace{2cm}
{\bf Abstract}
\end{center}

\begin{quote}
The two-point functions of the energy-momentum 
tensor and the Noether current are 
used to probe the $O(3)$ nonlinear sigma model in an energy range
below $10^4$ in units of the mass gap $m$. We argue that the 
form factor approach, with the form factor series trunctated at 
the 6-particle level, provides an almost exact solution of the model
in this energy range. The onset of the (2-loop) perturbative 
regime is found to occur only at energies around $100m$. 
\end{quote}
\vfill
\renewcommand{\thefootnote}{\arabic{footnote}}
\setcounter{footnote}{0}
\end{titlepage}
\newsection{Introduction}

The $O(3)$ Nonlinear Sigma (NLS) model has long been appreciated 
as a 2-dimensional testing ground for nonabelian gauge theories.
In particular it shares features like dynamical mass generation,
running coupling constant and (conjectured) asymptotic freedom
with its 4-dimensional cousins. A central issue is therefore to
determine the domain of validity of perturbation theory in the model. 
This requires an approach which is capable of describing the system 
both in the non-perturbative low energy and the 
perturbative high energy region. Although it could not have been
expected from the outset, the form factor approach \cite{KarWeisz, Smir}
(which essentially provides a low energy expansion of $n$-point 
functions based on the exact $S$-matrix \cite{ZZ}) turns out 
to be well suited for that purpose. Using the two-point functions
of the energy-momentum tensor and the Noether current to probe the system, 
we find that the form factor series truncated at the 6-particle level
provides an almost exact solution of the model in the energy range 
below $10^4$ in units of the mass gap $m$. 
A similar fast convergence of the low energy expansion has previously been
observed for the class of models with diagonal $S$-matrices 
\cite {res1,res2,res3} .
 
We compare our results with Monte Carlo 
data \cite{PSMC} and 2-loop perturbation
theory (PT). Within the range considered 2-loop PT appears to yield
an accurate (within one percent) 
description of the system only for energies above $100m$, 
provided one uses the known exact value 
of the Lambda parameter \cite{HMN} to fix its absolute normalization. 
In 4-dimensional gauge theories of course neither the exact 
Lambda parameter nor the (almost) exact form factor curve is 
available. Based on the low energy Monte Carlo data alone one 
might therefore be tempted to maximize the apparent domain of 
validity of PT by tuning the Lambda parameter such as to match 
the relevant part of the Monte Carlo data. Doing this in the 
NLS model would give a value for the Lambda parameter that is smaller
than the exact value by  about 10\%.
This emphasizes the importance to have an 
independent estimate for the onset of the (2-loop) perturbative 
regime. 
\newsection{Spectral representation of two point functions}

The form factors characterize an (integrable as well as non-integrable)
QFT in a similar way as the $n$-point functions do. Assuming
the existence of a resolution of the identity in terms of asymptotic
multi-particle states, the $n$-point functions can in principle be 
recovered from the form factors. In the important case of the two-point
functions this amounts to the well-known spectral representation. 
Explicitly for the Minkowski two-point function (Wightman function) 
of some local operator $\cO$ one obtains in a first step%
\be
W^{\cO}(x-y)= 
\sum_{n\geq 1}\frac{1}{n!}\int \prod_{j=1}^n \frac{d\th_j}{4\pi}\;
e^{-i(x^0-y^0)P_0^{(n)}(\th) - i(x^1 -y^1)P_1^{(n)}(\th)}\;
|f^{(\cO,n)}(\th)|^2 \;,
\ee  
where $f^{(\cO,n)}(\th) = \bra vac |\cO(0)|\th_n,\ldots,\th_1\ket$ are the 
form factors of $\cO$ and $P_{\mu}^{(n)}(\th) = \sum_i p_{\mu}(\th_i)$,
\newline $p_0(\th) = m\ch\th,\;p_1(\th) = m \sh \th$ are the eigenvalues of 
energy and momentum on an $n$-particle state.%
\footnote{Our kinematical conventions are: $(x^0,x^1)$ are coordinates
on 2-dimensional Minkowski space $\R^{1,1}$ with bilinear form  
$x\cdot y = x^{\mu}\eta_{\mu\nu}y^{\nu},\;\eta =\mbox{diag}(1,-1)$. 
Lightcone coordinates are $x^{\pm} = (x^0\pm x^1)/\sqrt{2}=x_{\mp}$. 
The normalization of the 1-particle states is 
$\bra\th_1|\th_2\ket =4\pi\,\delta(\th_1-\th_2)$, which corresponds to
the standard normalization in $d+1$ dimensions, specialized to $d=1$.
For simplicity we suppress internal indices for most of this section.}
The local operators are
classified by various quantum numbers, in particular by their Lorentz spin $s$
and their mass dimension $\Delta$. The form factors of a local operator with
quantum numbers $(\Delta,s),\;|s|\leq \Delta$ can be assumed to be of the 
form 
\be
f^{(n,\cO)}(\th) = i^s 
\left(\frac{m}{\sqrt{2}}\sum_j e^{\th_j}\right)^{\frac{\Delta + s}{2}}\; 
\left(\frac{m}{\sqrt{2}}\sum_j e^{-\th_j}\right)^{\frac{\Delta - s}{2}}\;
f^{(n)}(\th)\;,
\ee
where $f^{(n)}(\th)$ carries quantum numbers $(\Delta,s) =(0,0)$ and
is a function of the rapidity differences only. We shall henceforth 
always work with the `scalarized' form factors $f^{(n)}(\th)$ and drop 
the superscript `$\cO$'. 
For many purposes it is useful to rewrite (1) in the form of a 
K\"allen-Lehmann spectral representation. 
Changing integration variables according to 
\bas
&& u_i = \th_i -\th_{i+1}\;,\sspace 1\leq i\leq n-1\;,\sspace
\alpha = \ln\left(
\frac{m(e^{\th_1}+\ldots + e^{\th_n})}{M^{(n)}(u)} \right)\;,\nonum  
&& M^{(n)}(u)^2 = m^2[n + 2 \sum_{i<j}\ch(u_i+\ldots +u_{j-1})]\;
\eas    
and considering the case of an operator $\cO$ with quantum numbers
$(\Delta,s)$ one obtains
\ba
&& W^{\cO}(x-y) = -i\int_0^{\infty}d\mu\;\rho(\mu)\;
D^{(\Delta,s)}(x-y;\mu)\;,\nonum
&& \rho(\mu) = \sum_{n=1}^{[\mu/m]}\rho^{(n)}(\mu)\;,\sspace 
\rho^{(1)}(\mu) = \delta(\mu - m)\,|f^{(1)}|^2\;,\nonum
&& \rho^{(n)}(\mu) = \int_0^{\infty}
\frac{d u_1 \ldots d u_{n-1}}{(4 \pi)^{n-1}}\;|f^{(n)}(u)|^2\; 
\delta(\mu - M^{(n)}(u))\;,\;\;\;n\geq 2\;.
\ea
Notice that no problem of convergence arises for the spectral density.
First, each $n$-particle contribution exists because, for fixed $\mu$, 
the integrand has support only in a compact domain 
$V(\mu)\subset\R_+^{n-1}$ in which the form factors are bounded functions, 
so that the integration is well-defined.
Viewed as a function of $\mu$ one observes that $\rho^{(n)}(\mu)$ has 
support only for $\mu \geq m n$. 
Therefore only a finite number of terms (those with $n\leq [\mu/m]$, 
$[x]$ being the integer part of $x$) contribute to $\rho(\mu)$.  
Under some mild assumptions on the growth of $\rho(\mu)$ the existence 
of the spectral density then guarantees that of the 2-point function, as
defined through (3). The integration kernel $D^{(\Delta,s)}(x;m)$ in 
(3) is given by
\ba
D^{(\Delta,s)}(x;m) \is i\int\frac{d^2 p}{2\pi}\,\theta(p_0)
\delta(p^2 - m^2)\;(p_+)^{\Delta+s}(p_-)^{\Delta-s}\;
e^{-ip\cdot x} \nonum
\is (i\d_+)^{\Delta+s}\,(i\d_-)^{\Delta-s}\;D(x;m)\;,
\ea
where $D(x;m):=D^{(0,0)}(x;m)$ is the two-point function of a free scalar
field of mass $m$. (In general however $D^{(\Delta,s)}(x;m)$ does {\em not} 
coincide with the two-point function of some composite operator $\cO$ 
built from the scalar field). In the spinless case one obtains from the 
wave equation $D^{(\Delta,0)}(x;m)=(m^2/2)^{\Delta}\,D(x;m)$. The support
properties of the kernel functions can be extracted by evaluating them 
in terms of Bessel functions; in particular for strictly spacelike 
distances one finds
\be
D^{(\Delta,s)}(x;m) = i\left(\frac{m^2}{2}\right)^\Delta\,\frac{1}{2\pi}\,
\left(\frac{x^-}{x^+}\right)^s\;K_{2s}(m\sqrt{-x^2})\;,
\sspace x^2 <0\;.
\ee
Essentially this also yields the kernel function entering the spectral
representation of the Euclidean two-point function (Schwinger function).
The latter can be defined by $S^{\cO}(x_1,x_2) := W^{\cO}(-ix_2,x_1)$ 
for $x_2>0$ and then by analytic continuation to $x_2<0$. In the spectral  
representation (3) this yields $-i D^{(\Delta,s)}(-ix_2,x_1;m)$
for the Euclidean kernel function with the right hand of (5) taken for 
$D^{(\Delta,s)}(x;m)$. Notice that the spectral density is the same in 
the Minkowski and in the Euclidean case. Return now to the Minkowski space 
situation. For comparison with perturbation theory one needs the time-ordered
two-point function and its Fourier transform. Its spectral representation
is easily read off from (3), (4) 
\ba
G^{\cO}(x-y) \is -i\int_0^{\infty}d\mu\;\rho(\mu)\;
G^{(\Delta,s)}(x-y;\mu)\;,\nonum
G^{(\Delta,s)}(x;m)\is \theta(x_0)\,D^{(\Delta,s)}(x;m) - 
\theta(-x_0)\,[D^{(\Delta,s)}(x;m)]^* \nonum
\is  \int\frac{d^2 p}{(2\pi)^2}\,e^{-ip\cdot x}\;
\frac{(p_+)^{\Delta+s}(p_-)^{\Delta-s}}{p^2 - m^2 + i\epsilon}
\ea
For the Fourier transform this means
\be
\widetilde{G}^{\cO}(p) =
i \left(-\frac{1}{2}\right)^{\Delta}\,\left(\frac{p_+}{p_-}\right)^s 
\;I(-p^2)\;,\sspace I(z) = z^{\Delta} 
\int_0^{\infty}d\mu\frac{\rho(\mu)}{z+\mu^2 -i\epsilon}\;.
\ee
The definition of $I(z)$ was chosen such that it has a cut along the 
negative real axis and one can recover the spectral density from
the discontinuity along this cut.

As an example consider the case of the energy-momentum (EM) tensor. 
It is of additional interest, because its spectral density is closely
related to the Zamolodchikov C-function \cite{ZC}. 
The spectral representation of 
the Minkowski 2-point function is 
\bas
&& W^{EM}_{\mu\nu,\rho\sigma}(x-y) = -i\int_0^{\infty}d\mu\;\rho(\mu)\;
D_{\mu\nu,\rho\sigma}(x-y;\mu)\;,\nonum
&& D_{\mu\nu,\rho\sigma}(x;m) = i\int\frac{d^2 p}{2\pi}\,\theta(p_0)
\delta(p^2 - m^2)\;e^{-ip\cdot x}\;(\eta_{\mu\nu}p^2 - p_{\mu}p_{\nu})
(\eta_{\rho\sigma}p^2 - p_{\rho}p_{\sigma})\;.
\eas
The lightcone components correspond to the $SO(1,1)$-irreducible pieces;
in particular $D_{++,++}(x;m)$ and $D_{+-,+-}(x;m)$ coincide with
$D^{(2,2)}(x;m)$ and $D^{(2,0)}(x;m)$, respectively. Combining (3) and
(5) one finds for the behaviour at small spacelike distances
\ba
W^{EM}_{++,++}(x) \is \left(\frac{x^-}{x^+}\right)^2\left[
\frac{c}{2\pi^2}\;\frac{1}{(x^2)^2} + O\left(\frac{1}{x^2}\right)\right]\;,
\sspace x^2 <0\;,\nonum
c\is 12 \pi \int_0^{\infty}d\mu \,\rho(\mu)\;.
\ea
The number $c$ as defined here coincides with the central charge
of the Virasoro algebra in the conformal field theory describing the 
UV fixed point of the renormalization group. This latter fact is part
of the statement of Zamolodchikov's C-theorem.%
\footnote{In particular c is finite; for other operators the integral 
over the spectral density will in general not converge.}
The proof of the C-theorem is particularly transparent from the viewpoint 
of the spectral representation \cite{CFL}. The normalization is such that
$H = \int_{-\infty}^{\infty}dx^1\,T_{00}(x)$ has eigenvalue 
$\sqrt{p^2 + m^2}$ on an asymptotic single particle state, where 
$m$ is the mass gap. For the 2-particle form factor of the $T_{+-}$
component this implies
\be
f^{(T_{+-},2)}(\th + i\pi, \th) = m^2\;,
\ee
which fixes the normalization of all higher particle EM form factors.
The scalarized form factors of all components can be interpreted as the 
form factors of a single field $\tau$ defined by $T_{\mu\nu} = 
\epsilon_{\mu\rho}\epsilon_{\nu\sigma}\partial^{\rho}\partial^{\sigma}\tau$.
The conservation equation is then built into the parametrization (2).

Also a Noether current comes with an intrinsic normalization, which 
arises from the Lie algebra of its conserved charges. In the case of 
the $O(3)$ NLS model one has $[Q_a\,,Q_b] = i\epsilon_{abc}\, Q_c$ and
$$
Q_a\,|\th\ket_b = i\epsilon_{abc}\,|\th\ket_c\;,
$$
where $Q_a= \int dx^1 J_{0,a}(x)$ is the conserved charge of the 
$O(3)$ Noether current. For the 2-particle form factor of $J_{0,a}$
this implies
\be
f^{J_{0,a}}_{bc}(\th +i\pi, \th) = -2 i\,m\,\epsilon_{abc} \ch\th\;,
\ee
which fixes the normalization of all higher current form factors.
The scalarized form factors of both (spacetime) components can be 
interpreted as the form factors of the field $\tau_a$ defined through 
$J_{\mu,a} = \epsilon_{\mu}^{\;\;\nu}\,\d_{\nu}\tau_a$. The two point 
function of $J_{\mu,a}(x)$ and $J_{\nu,b}(y)$ and hence the spectral 
density is of course proportional to $\delta_{ab}$. 

\newsection{Polynomial $O(3)$-irreducible form factors}

Solutions of the form factor equations with the $O(3)$-invariant $S$-matrix 
\cite{ZZ} have been obtained by Kirillov and Smirnov \cite{KirSmir} by a 
fusing procedure from the previously known form factors of the $SU(2)$ 
Thirring model \cite{Smir}. Unfortunately these solutions refer to a basis 
in the tensor product ${\bf 3}^{\otimes n}$ (${\bf 2 l+1}, l\in \N$
being the $O(3)$ irreps of spin $l$) that is related to the canonical 
basis through a complicated rapidity-dependent basis transformation. 
Since the most interesting local operators in the model all 
transform irreducibly under $O(3)$ one has to determine the intertwiners
$$
Q_l^{(n)}:{\bf 3}^{\otimes n} \rra {\bf 2l+1}\;
$$
to the canonical basis in $\C^{2l+1} \cong {\bf 2l+1}$. 
If one starts from the Kirillov-Smirnov basis in ${\bf 3}^{\otimes n}$ 
also these intertwiners will be rapidity-dependent and their explicit 
computation is almost as difficult as the computation of the form factors 
themselves. We therefore found it easier to start afresh and to work with 
the canonical basis in ${\bf 3}^{\otimes n}$ and constant intertwiners.  
A detailed exposition of the technique and the results will be given 
elsewhere \cite{BN}. 

In survey we studied the following four local operators,
which seem to be the most interesting ones in the $O(3)$ model:
The `fundamental' spin field $S$, the Noether current, 
the EM tensor and the topological charge (TC) density.
For orientation we tabulate the expressions in terms of the field 
$S$, their quantum numbers $(l,\Delta)$ and the particle numbers of the 
form factors computed.  
\vspace{5mm}

\hspace{1cm}
\begin{tabular}{c|c|c|c}
$\cO$   &   $\cO[S]$        &    $(l,\Delta)$   & ff's computed  \\[0.5ex] 
\hline 
spin    &   $\;\;S\;\;$     &    $\;\;(1,0)\;\;$ &  $\;\;1,3,(5)$ \\
current &   $\;\;j_{\mu} = S\times\partial_{\mu}S\;\;$ 
        & $\;\;(1,1)\;\;$   & $\;\;2,4,(6)\;\;$                   \\
EM tensor & $\;\;T_{\mu\nu} = \d_{\mu}S\cdot\d_{\nu}S -\frac{1}{2} 
            \eta_{\mu\nu}\,\d^{\rho}S\cdot\d_{\rho}S\;\;$ 
        & $\;\;(0,2)\;\;$   & $\;\;2,4,(6)\;\;$              \\
TC density & $\;\;q = S\cdot(\d_{\mu}S\times \d_{\nu}S)\,
                      \epsilon_{\mu\nu}\;\;$ 
        & $\;\;(0,2)\;\;$   & $\;\;3,(5)\;\;$ 
\end{tabular}
\vspace{5mm}

The explicit formulae can be found in \cite{BN}. The brackets around 
the last particle number indicate that the resulting expressions for 
the form factors are too long to be communicated in print; instead we 
shall give the expressions for their modulus squares, which is also 
the quantitiy entering the spectral densities.%
\footnote{The difference in size is enormous. The 6-particle form factor
of the current has about 3.4 Mbytes, i.e. approx. 700 A4 pages; the 
most compact form of the square requires less than one page.}
Higher particle contributions could be computed in principle, although
more refined techniques would be necessarry to circumvent limitations 
in computer power. It seems however that higher particle contributions
to the spectral density are strongly suppressed in the $O(3)$-model. 
Indeed, in section 4 we shall argue that truncating the form factor 
series at the 5 or 6 particle level provides an `almost exact' solution 
of the model up to energies of about $10^4$ times the mass gap. 
`Almost exact' means that within this energy range the deviation of the 
true spectral densities from the truncated ones
is estimated to be less than a percent. If one accepts the conventional
wisdom about the asymptotic freedom of the model at very high energies,
the combination of the non-perturbative form factor results 
(for energies up to $10^4$ in units of the mass gap) and the
perturbative expansion at energies above $10^4$ would provide a 
solution of the model sufficient for most practical purposes.  
For issues of principle it would nevertheless be highly desirable to 
get some control on the higher particle contributions,
in particular because the `conventional wisdom' has been 
challenged \cite{PS}. Some first results in this direction will also be 
presented in \cite{BN}. An important such issue is trying to actually 
prove (or disprove) asymptotic freedom. 

We shall not discuss the construction of the form factors here. 
Let us just mention the decomposition into $O(3)$ irreducible 
components. Let  $f_{l,A}(\th)$ denote the component of $f_A(\th)$ 
in ${\bf 2l+1}$. Explicitly $f_{l,A}(\th)$ is an $n$-index $O(3)$ 
tensor parametrized by $m_l(n)$ independent functions, where $m_l(n)$
is the multiplicity with which ${\bf 2l+1}$ occurs in 
${\bf 3}^{\otimes n}$. The choice of a parametrization 
amounts to a choice of an intertwiner $Q_l$ between the $O(3)$ 
representations ${\bf 3}^{\otimes n}$ and ${\bf 2l+1}$. 
Picking a basis in both spaces one can write 
\be
f_{l,A}(\th)= Q_{l,A}^{\alpha} \,g_{\alpha}(\th)\;.
\ee
As remarked before, in contrast to \cite{KirSmir}, we work with the 
canonical basis in ${\bf 3}^{\otimes n}$ and constant 
(rapidity-independent) intertwiners $Q_{l,A}$. (This 
means that each component $Q_{l,A}^{\alpha}$ can explicitly be written as a 
combination of antisymmetric epsilon tensors and Kronecker deltas.) 
For the components $g_{\alpha}(\th)$ we used the following Ansatz
\ba
&& g_{\alpha}(\th) = \gbar_{\alpha}(\th)\,\;\pi^{3(n-2)/2} 
\prod_{k>r}\psi(\th_k -\th_r)\;,\sspace 1\leq \alpha \leq m_l(n)\;,\nonum
&& \psi(\th)=\frac{\th -i\pi}{\th(2\pi i -\th)}\tanh^2(\th/2)\;.
\ea
The main structural result can then be summarized as follows \cite{B,BH,BN}:

{\bf Fact:} The form factor equations for the $O(3)$ NLS model 
decompose into decoupled recursive systems for the $O(3)$-irreducible
components $g^{(n)}(\th) =(g_1,\ldots, g_{m_l(n)})(\th)$. 
There exist unique sequences of solutions $(g^{(n)}(\th))_{n\geq 1}$
of the form (12) s.t. $\gbar^{(n)}(\th) =
(\gbar_1,\ldots, \gbar_{m_l(n)})(\th)$ are polynomials in the 
rapidities of total degree $ N= \frac{1}{2}(n^2 - 3 n) + N_0$ and 
partial degree $p= n-2 +p_0$.

Here `partial degree' means the degree in an individual variable.
For the operators considered before the initial values are given 
by $(N_0,p_0)=(0,-1)$ for EM tensor \& TC density and 
$(N_0,p_0) = (1,0)$ for Current \& Spin. The fact that the form 
factors can be reduced to polynomial expressions is specific for 
the $O(3)$ model and will cease to hold for the $O(N)$ NLS models 
with $N>3$. The first part of the above statement in particular means 
that the $O(3)$ spin is a good quantum number for form factor 
{\em sequences}: If one member of the sequence is 
$O(3)$-irreducible of spin $l$ then all other members will be too. 
This is only seemingly trivial; it will not hold for any Lie group 
and any $S$-matrix. 

For the computation of the spectral densities and the two point 
functions the modulus square of the form factors is needed.
A version of $|f_l^{(n)}(\th)|^2$ that allows analytic continuation to 
complex rapidities is 
\ba
&& f_{l,a_n\ldots a_1}(\th_n,\ldots,\th_1)
f_{l,a_1\ldots a_n}(\th_1,\ldots,\th_n) =
g_{\alpha}(\th_n,\ldots,\th_1)\, C_{\alpha,\beta}\, 
g_{\beta}(\th_1,\ldots,\th_n) \nonum
&& \sspace = C_l\,\frac{1}{4} \pi^{3n-2}\;\Gbar_l^{(n)}(\th) 
\prod_{k>r}\psi(\th_k - \th_r)
\prod_{i<j}\psi(\th_i - \th_j)\;,
\ea
where $C_{\alpha,\beta}$ is a constant `metric' and $l=0,1$ correspond to
the EM tensor \& TC density and Current \& Spin series, respectively. 
The prefactor $C_l$ is given by $C_0 =3$ and $C_1 =\delta_{ab}$.
The so defined square is a completely symmetric function
of the rapidities; in particular its polynomial part $\Gbar^{(n)}_l(\th)$  
is a completely symmetric, boost invariant  
polynomial in the rapidities. (It depends only on the rapidity
differences.)
By choosing
an appropriate basis in the space of symmetric polynomials it 
can be brought into a very compact form, as compared with
what one might guess from the size of the form factors
(c.f. footnote 3). A convenient basis is obtained as follows:
For fixed $n$ let $\sigma_1^{(n)},\ldots,\sigma_n^{(n)}$  
denote the elementary symmetric polynomials in $\th_1,\ldots,\th_n$, i.e.
$$
\sigma_k^{(n)} =\sum_{i_1<\ldots <i_k}\th_{i_1}\ldots \th_{i_k}\;.
$$
Then define polynomials $\tau^{(n)}_k$ by
\ba
&& \tau^{(n)}_k(\th_1,\ldots,\th_n) = 
\sigma^{(n)}_k(\widehat{\th}_1,\ldots,\widehat{\th}_n)\;,
\sspace 2\leq k\leq n\;,\nonum
&& \tau^{(n)}_1 = \frac{1}{n}\sigma^{(n)}_1\;,\sspace
\widehat{\th}_j = \th_j -\frac{1}{n}(\th_1 + \ldots +\th_n)\;.
\ea
All monomials in the $\tau^{(n)}_k,\;k\geq 2$ are manifestly boost 
invariant. (The price to pay is that the partial degree is no longer
manifest.) For the sake of illustration we display the polynomial part 
of the 4-particle squares of the EM tensor and the Current in this 
basis (where the superscripts `$(4)$' are suppressed):
\begin{eqnarray}
&&\mbox{EM tensor:}\nonum 
&& \Gbar_0^{(4)}(\th) = 4(\tau_2^2 +12 \tau_4) - \pi^2 32\tau_2 + 28 \pi^4\;,
\nonum
&&\mbox{Current:}\nonum 
&& \Gbar_1^{(4)}(\th)
 = -4(6\tau_2^3 + 9\tau_3^2 + 40 \tau_2\tau_4) + 
   8 \pi^2(25 \tau_2^2 + 44\tau_4) - 448 \pi^4 \tau_2 + 272 \pi^6\;.
\end{eqnarray}

\newsection{Results for spectral densities and two-point functions}

\setcounter{totalnumber}{2}
\renewcommand{\textfraction}{0.01}
\renewcommand{\bottomfraction}{0.99}
\floatsep 0mm

Knowing the form factor squares the evaluation of the spectral
densities and the two-point functions is in principle straightforward.
The integrations in (3) can be done numerically to good accuracy.
Here we shall restrict attention to the EM tensor and the Noether 
current. For the spectral densities an accuracy of $10^{-3}$ was used and 
the results for the EM tensor and the Noether current are shown in figures 1
and 2, respectively. In the EM case one can use equation (8) to 
compute the $n$-particle contributions to the central charge. One
finds
\be
c^{(2)} = 1.603\;,\;\;c^{(4)} =0.194\;,\;\;c^{(6)} = 0.072\;,\sspace
c^{(n)} = 12\pi \int_0^{\infty}d\mu \,\rho^{(n)}(\mu)\;.
\ee
If one accepts asymptotic freedom as a working hypthesis, (tree level)
PT predicts $c=2$, corresponding to the two unconstrained bosonic
degrees of freedom. The form factor computation shows that this is 
compatible with the non-perturbative low energy dynamics of the model
and that even for an extreme UV quantity like the central charge
the low particle contributions dominate. An alternative non-perturbative
consistency check is provided by the thermodynamic Bethe Ansatz,
which also yields $c=2$ \cite{FZ}. Let us point out that the value $c=2$ does 
not carry much information about the nature of the UV limiting CFT. 
For a number of reasons it cannot be that of two decoupled free bosons.

\begin{figure}[htb]
\begin{center}
\leavevmode
\epsfxsize=170mm
\epsfbox{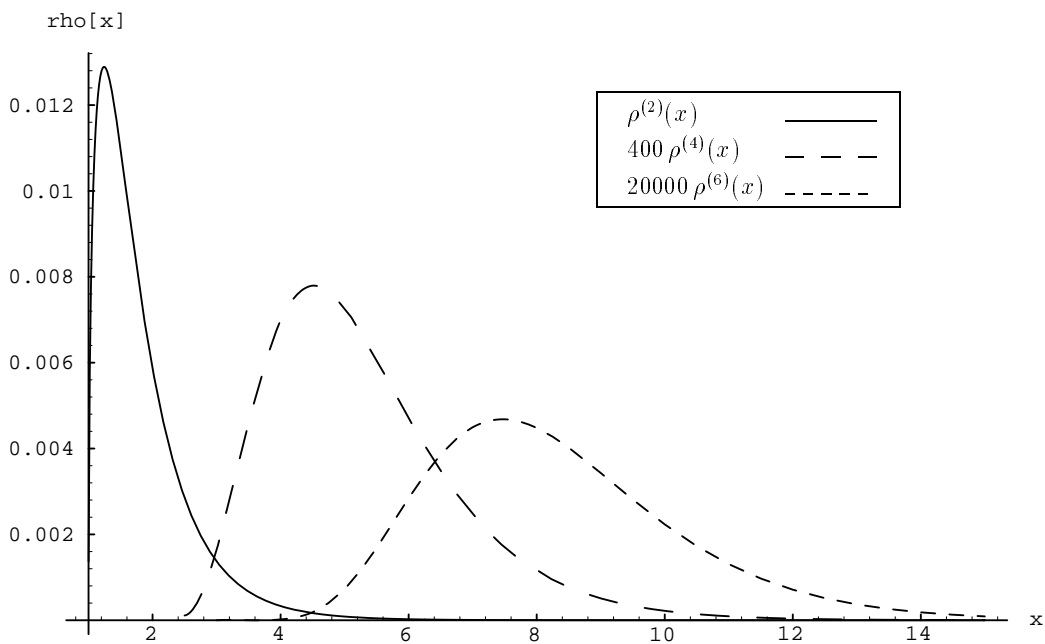}
\end{center}
\caption{2-, 4-, and 6-particle contribution to the spectral density
of the EM tensor as a function of $x= \log_2(\mu/m)$.}
\label{spec1}
\end{figure}
\begin{figure}[htb]
\begin{center}
\leavevmode
\epsfxsize=170mm
\epsfbox{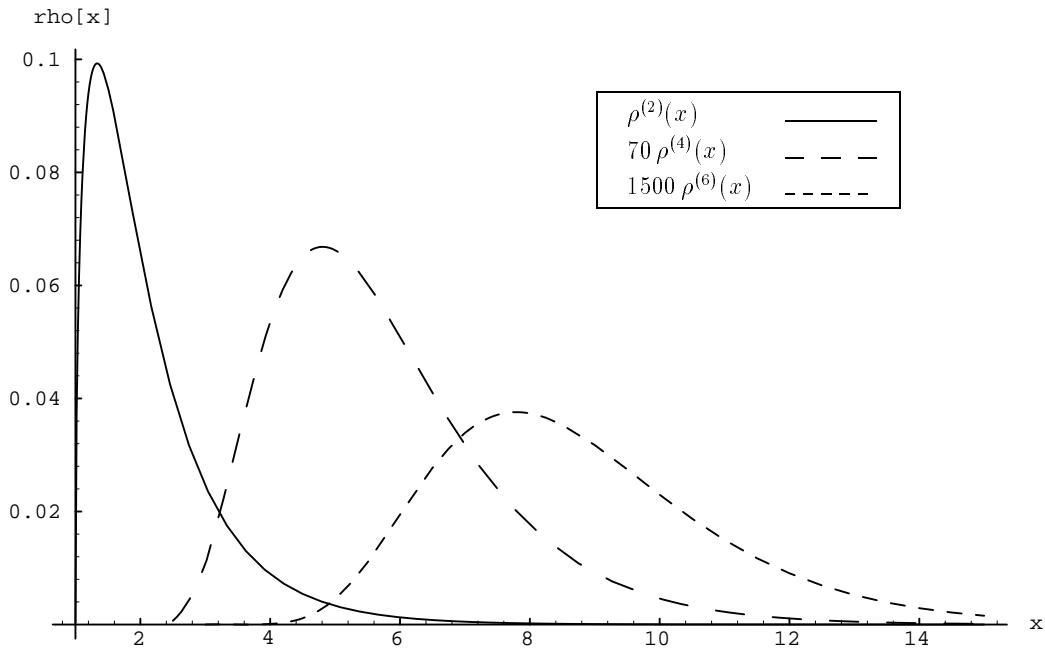}
\end{center}
\caption{2-, 4-, and 6-particle contribution to the spectral density
of the current as a function of $x= \log_2(\mu/m)$.}
\label{spec2}
\end{figure}
Figures 1 and 2 illustrate some general features of the $n$-particle
contributions to the spectral densities. First one observes a strong 
suppression of the higher particle contributions, as far as the value 
of the maximum and the enclosed area are concerned. Generally 
speaking the maximum of $\rho^{(n)}(\mu)$ is smaller by 1.5 to 2.5 
orders of magnitudes compared to the maximum of $\rho^{(n-2)}(\mu)$, while 
the position of the maximum is shifted towards higher energies. 
Nevertheless at sufficiently high energies the $n$-particle contribution
overtakes the $(n-2)$-particle contribution, i.e. $\rho^{(n)}(\mu)>
\rho^{(n-2)}(\mu)$, for $\mu > \mu_n$, where $\rho^{(n)}(\mu_n)=
\rho^{(n-2)}(\mu_n)$. (This feature is not directly visible in figures 1 
and 2 because of the relative rescaling and the insufficient magnification 
of the intersection region.) 
The point of intersection $\mu_n$ is of particular interest
because it provides an {\em intrinsic} measure for the quality of the 
approximation made by truncating the form factor series at the 
$n$-particle term: Since $\mu_{n+2} \gg \mu_n$ the $(n+2)$-particle
contribution can safely be ignored up to energies $\mu \lsim \mu_n$    
in the sense that the correction to $\rho^{(2)}(\mu) +\ldots + 
\rho^{(n)}(\mu)$ would be at most (of the order of) a percent. The 
results for these points of intersection $\big(\mu_n,\rho^{(n-2)}(\mu_n) =
\rho^{(n)}(\mu_n)\big)$ are as follows:
\ba
\mbox{EM:}\;\;&\mbox{$n=4$:}\;\;\;(1.7\cdot 10^2, 4.7\cdot 10^{-6})\;,\;\;
&\mbox{$n=6$:}\;\;\;(4.5\cdot 10^5, 2.5\cdot 10^{-10})\;,\nonum
\mbox{Current:}\;\;&\mbox{$n=4$:}\;\;\;(1.6\cdot 10^2,3.8\cdot10^{-4})\;,\;\;
&\mbox{$n=6$:}\;\;\;(1.2\cdot 10^6, 3.2\cdot 10^{-8})\;.
\ea
Thus, the form factor series truncated at 6 particles should provide 
accurate results for the spectral density up to energies $O(10^5)$ --
results, which can then be compared with those obtained by other 
techniques. 

In the high energy regime one expects the onset
of asymptotic freedom and the perturbative predictions should coincide 
fairly well with the form factor curves. This is indeed the case as can
be seen in figure 3. What is shown is the Fourier transform $I(p^2)$ of 
the 2-point function of the Noether current, computed once in 2-loop PT 
and once via (7) by truncation of the form factor series. 
In performing the integral transformation (7) one can see that accurate 
results for $\rho(\mu)$ in the region $2 \lsim \mu \lsim 10^5$ produce 
accurate results for $I(z)$ in the range $z\lsim 10^4$. In the energy range 
$100 \lsim \mu/m \lsim 10^4$ the coincidence of the 2-loop PT curve
and the $2+4+6$-particle form factor curve indicates that the system 
is desribed well by 2-loop PT in this regime. The result (17) for the 
intersection point $\mu_6$ suggests that the deviation at yet larger 
energies can entirely be attributed to the truncation of the form factor
series. Notice however that the onset of the (2-loop) perturbative 
regime occurs at much higher energies $\mu/m \gsim 100$ than is 
sometimes pretended in the 4-dimensional counterpart of this situation. 
Figure 4 is a magnification of the low energy region of figure 3, 
where non-perturbative effects are expected to become important. 

\begin{figure}[htb]
\leavevmode
\vskip 5mm
\epsfxsize=170mm
\epsfbox{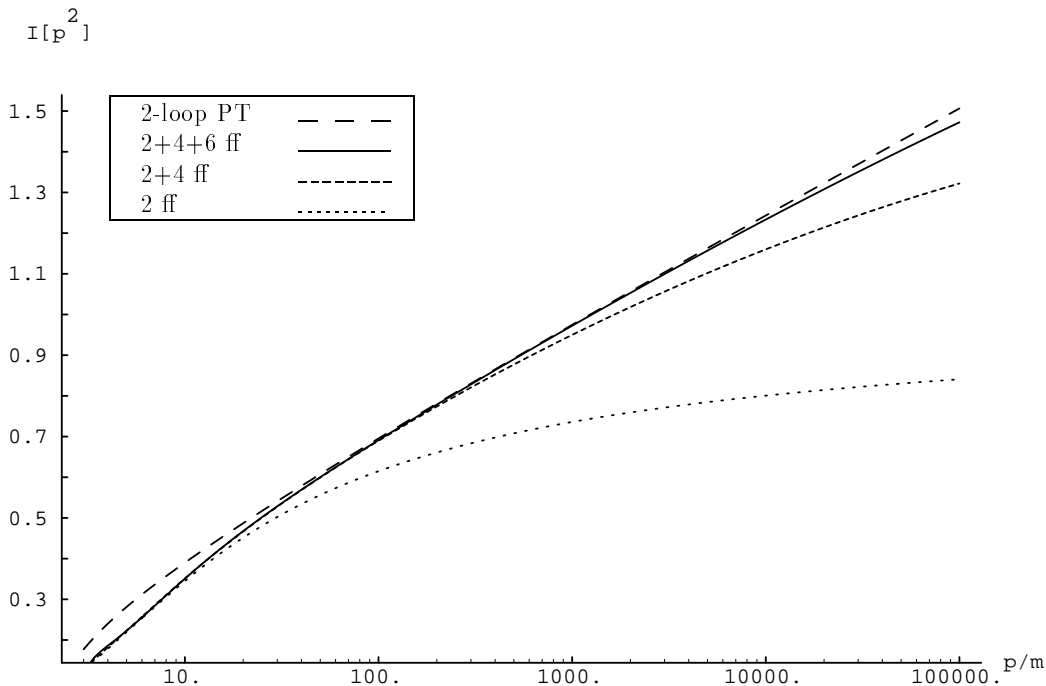}

\caption{Comparison: Form factor approach versus 2-loop
perturbation theory; logplot of $I(p^2)$ against $p/m$.} 
\end{figure}
\begin{figure}[htb]
\leavevmode
\epsfxsize=170mm
\epsfbox{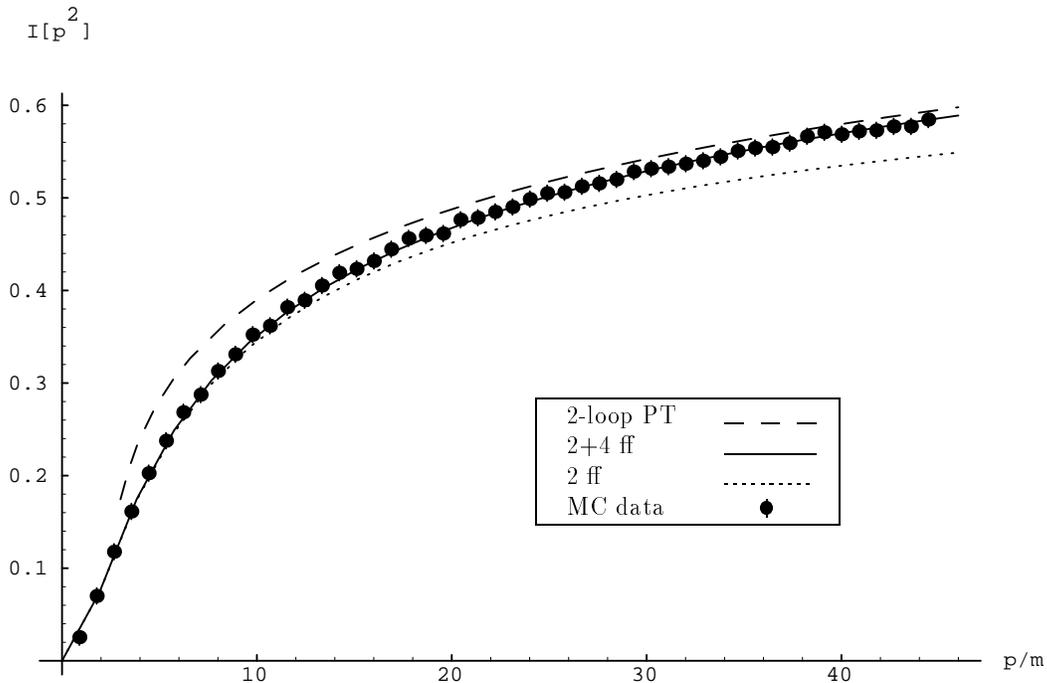}

\caption{Low energy region of figure 3. Comparison: Form factor approach, 
Monte Carlo data and 2-loop perturbation theory; 
$I(p^2)$ plotted against $p/m$.} 
\end{figure}
Here
Monte Carlo simulations provide an alternative non-perturbative technique 
to probe the system \cite{LW,PSMC}. The simulations were made \cite{PSMC} 
using a Wolff-type cluster algorithm on a 460 square lattice at inverse 
coupling $\beta = 1.80$ (correlation length $\xi = 65.05$). The agreement 
between the MC data and the form factor results is excellent. One also 
sees that for energies between 30 and 45 the PT curve runs almost parallel 
to the MC data. Without the guidance of the form factor result one would 
thus be tempted to match both curves by tuning the Lambda parameter 
approriately. Doing this however, the Lambda parameter comes out wrong by 
about 10\% (from below), as compared with the known 
exact result \cite{HMN}. 
Generally speaking one sees that a determination of the Lambda parameter
from MC data and (2-loop) PT 
about the onset of the (2-loop) perturbative regime enters. 
The form factor results in the $O(3)$-NLS 
model show that this regime sets in only at relatively high 
energies $\mu/m \gsim 100$, which should at least be taken as a warning 
in 4-dimensional gauge theories.
\vspace{5mm}

{\tt Acknowledgements:} We wish to thank H. Lehmann, A. Patriasciou, 
E. Seiler and P. Weisz for stimulating discussions. A. P. and E. S. we 
thank in addition for generously allowing us to use their data prior to
publication. M. N. acknowledges support by the Reimar 
L\"ust fellowship of 
the Max Planck Society. T. H. acknowledges support by the D.O.E.
(cooperative research agreement DE-FCO2-94ER40818).

\newpage
%

\end{document}